\documentclass[pra,showpacs,twocolumn,superscriptaddress,floatfix]{revtex4-1}

\usepackage[dvipsnames]{color}
\usepackage{graphicx}
\usepackage{bm}
\usepackage{amssymb}
\usepackage{amsmath}
\usepackage[normalem]{ulem}

\newcommand{\be}{\begin{equation}}
\newcommand{\ee}{\end{equation}}
\newcommand{\bea}{\begin{eqnarray}}
\newcommand{\eea}{\end{eqnarray}}

\newcommand{\la}{\langle}
\newcommand{\ra}{\rangle}

\newcommand{\mbf}[1]{\mathbf{#1}}
\newcommand{\abs}[1]{\left|#1\right|}
\begin{document}
\title{Testing density-functional approximations on a lattice and the limits of the related Hohenberg-Kohn-type theorem}
\author{Vivian Fran\c{c}a}
\affiliation{Institute of Chemistry, S\~{a}o Paulo State University, Brazil}
\author{Jeremy P. Coe}
\affiliation{Institute of Chemical Sciences, School of Engineering and Physical Sciences, Heriot-Watt University, Edinburgh, EH14 4AS, United Kingdom.}
\author{Irene D'Amico}
\affiliation{Department of Physics, University of York, York YO10 5DD, United Kingdom}
\affiliation{Instituto de Fisica de S{\~a}o Carlos, Universidade de S\~ao Paulo, 13560-970, S\~ao Carlos, S\~ao Paulo, Brazil}

\begin{abstract}
We present a metric-space approach to quantify the performance of density-functional approximations for interacting many-body systems and to explore the validity of the Hohenberg-Kohn-type theorem on fermionic lattices. This theorem demonstrates the existence of one-to-one mappings between particle densities, wave functions and external potentials. We then focus on these quantities, and quantify how far apart in metric space the approximated and exact ones are.
We apply our method to the one-dimensional  Hubbard model for different types of external potentials,  and assess its validity on one of the most used approximations in density-functional theory, the local density approximation (LDA). We find that the potential distance may have a very different behaviour from the density and wave function distances, in some cases even providing the wrong assessments of the LDA performance trends. We attribute this to the systems reaching behaviours which are borderline for the applicability of the one-to-one correspondence between density and external potential. On the contrary the wave function and density distances behave similarly  and are always sensitive to system variations.  Our metric-based method correctly predicts the regimes where  the LDA performs fairly well and the regimes where it fails. This suggests that our method could be a  practical tool for testing the efficiency of density-functional approximations.

\end{abstract}
 \maketitle

\section{Introduction}

The description and understanding of materials, nano-structures, atoms, molecules, and of their properties  is clearly non trivial, as these are interacting and inhomogeneous many-body systems, and their main variable in usual quantum approaches, the wave function, is a $3N$-dimensional function, $N$ being the total number of particles. 
In this context density-functional theory (DFT) \cite{HK} is a powerful alternative method, and uses as its main variable, the particle density, a function of only $3$ dimensions. This method is based on mapping the interacting many-body system into a fictitious non-interacting one, the Kohn-Sham (KS) system~\cite{KS}, whose potential is constructed to give the same density of the original interacting system, but can be solved much more easily. The Hohenberg-Kohn theorem then ensures that, given the exact density, any other property of the interacting system could be, in principle, calculated.
Although DFT is in principle an exact theory, in practice one has to make use of approximations in order to obtain the KS potential, and thus the particle density and any other desired property, as a density functional. Hence calculations within DFT provide approximate results, whose accuracy depends on the quality of the approximations used for the density functionals.
The typical way to check accuracy and optimize density functionals' approximations has been to try to reproduce ground state total energies.  Recently, however, a critique of this method have been raised \cite{Science2017}, where indeed a call to the community for optimization methods based on more physical quantities than just the total system energies has been put forward.

Lattice models, such as the Heisenberg or Hubbard models \cite{HeisenbergReview,OrigHubbard} have been extremely important for the understanding of strongly-correlated many-particle systems. Despite their simplicity,
when employed appropriately, these models are able to capture phenomena sufficiently accurately:  the Hubbard model has been shown to reproduce the Mott metal-insulator transition, and has been recently associated to the behaviour of `exotic' systems such as  inhomogeneous superfluidity in spin-imbalanced systems \cite{fflo1,fflo2}, chains of Bose-Einstein condensates \cite{HubbardBoseEinstein16}, or entanglement in nanostructures \cite{HubbardNano10,EPL2011}.  However, when  interactions and inhomogeneities are included, even these lattice models can rapidly become computationally intractable as the size of the systems increase.  This is where the powerful concept of DFT on a lattice \cite{LDFT1,LDFT2,LDFT3} becomes useful.  By using approximations within the approach of lattice DFT, specifically the local density approximation (LDA) \cite{LimaPRL2003,fvc,Capelle13},  the constraint on the maximum system size can be considerably loosened, and larger lattice systems can be accurately modelled.
DFT on a lattice is built on foundations \cite{PhysRevB.52.2504,WU1,Godby95} similar to the original Hohenberg-Kohn theorem for standard DFT. However these foundations were completed only recently \cite{EPL:2015}  when the last part of the Hohenberg-Kohn-type theorem $-$ the one-to-one mapping between ground-state density and external potential, and ground-state wave function and external potential $-$ was demonstrated, and their limitations discussed.

In this context, we evaluate the use of specifically designed metrics for density, wave function and potential to appraise lattice-DFT approximations beyond total energy arguments, and use the same tools to explore the limits of the Hohenberg-Kohn-type theorem on a lattice. Our results show that indeed these two subjects are intimately connected.  The method we propose applies to any fermionic single-band lattice system and could be used on any DFT approximation. Here we will concentrate on the LDA. As practical $-$ and exactly solvable $-$ test-bed examples, we will demonstrate our method on the one-dimensional Hubbard model considering short to medium size chains of varying particle number, interaction strength, and applied external potentials.

\section{Theoretical and computational methods}
DFT demonstrates that, for time-independent systems with a spin-independent external potential, ground-state particle density, the corresponding wave function, and the external potential are sufficient to describe a quantum system. We wish then to rigorously appraise DFT approximations by calculating how well they reproduce the exact values of these three quantities, and we will do so by using metric spaces.

We note that, to this aim, we cannot simply compare exact and Kohn-Sham systems, as nor the wave function, neither the external potential of the KS system are constructed to reproduce the exact ones. We then follow the route proposed in \cite{Coe08PRB,CoePRA09}: we will consider, together with the exact, that unique interacting system constructed to
have the same density as the non-interacting Kohn-Sham system obtained by using the DFT approximation, and with the same particle-particle interaction operator as the exact interacting system. We will construct it via reverse engineering, by using the iterative scheme in Ref.~\cite{EPL:2015} to find the potential of the interacting lattice system given the approximated density. At this point, the distance between {\em all} quantities of interest can be assessed by using appropriate metrics.

In this paper, to illustrate our method, we focus on one of the most used density-functional approximations, the LDA; we will introduce below all the elements of the method outlined above.

\subsection{The interacting-LDA system}
The  ``interacting-LDA'' system, the i-LDA  introduced in Ref.~\cite{Coe08PRB}, is the uniquely defined many-particle {\it interacting} system with the same many-body interaction operator as the original exact Hamiltonian , but whose ground-state density is the LDA density which is found by solving the Kohn-Sham equations within the LDA for the original many-body system.

The advantage of using the  i-LDA system is that its many-body interacting wave function tends to the exact many-body wave function for {\it all} interaction regimes for which the LDA is a good approximation. This is a fundamental difference with the corresponding KS-LDA wave function which is by definition non-interacting, and therefore it is  able to reproduce the exact many-body wave function only in the non-interacting limit. In addition, because the many-body interaction operator is the same for the exact and the i-LDA Hamiltonians, then the better the LDA performance, the closer the corresponding two external potentials $v$ and $v_{i-LDA}$ should be.

The properties of the exact, KS-LDA, and i-LDA systems are summarized in Table \ref{table1}.

\hspace{2cm}
\begin{table}[h!]
\begin{tabular}{|c|c|c|c|}
  \hline
  % after \\: \hline or \cline{col1-col2} \cline{col3-col4} ...
  ~ & density & wave function & external potential \\ \hline
  exact & $n$ & interacting $\Psi$ & $v$ \\
  KS-LDA & $n_{LDA}$ & non-interacting $\phi_{KS-LDA}$ & $v_{KS-LDA}$ \\
  i-LDA & $n_{LDA}$ & interacting $\Psi_{i-LDA}$ &  $v_{i-LDA}$ \\ \hline
\end{tabular}
  \caption{Properties of the exact, KS-LDA, and i-LDA systems. }\label{table1}
  \end{table}

\subsection{Iterative scheme on a lattice}
Consider a generic lattice Hamiltonian with an external potential  $v_j$ at site $j$, and so described by
\begin{equation}
\hat{H}=\hat{H}_{0}+\sum_{j}v_j\hat{n}_{j}. \label{H}
\end{equation}
Here $\hat{n}_{j}=\hat{c}^\dag_j\hat{c}_j$ the site $j$ occupation operator and $\hat{c}^\dag_j$ ($\hat{c}_j$) the fermionic creation (annihilation) operators.
The ground state wave function for this system is $|\Psi\ra$, such that

\begin{equation}
\la\Psi| \hat{n}_{i}H |\Psi\ra=En_{i}
\end{equation}
where $E$ is the ground state energy and we have used $n_{i}=\la\Psi |\hat{n}_{i}|\Psi\ra$.
We may then use the iterative scheme of Ref.~\cite{EPL:2015}
to find the potential which gives the density $n_{i}^{target}$ from an initial trial potential $v_{i}^{(0)}$.
The recursive formula is
\begin{equation}
 v_{i}^{(k+1)}=\frac{\left(n_{i}^{(k)}-n_{i}^{target}\right)|E^{(k)}|}{\la\Psi^{(k)} |\hat{n}_{i}^{2}|\Psi^{(k)}\ra} +v_{i}^{(k)}.
\label{eq:schemeHub}
\end{equation}
At convergency, $v_{i}^{(k+1)}=v_{i}^{(k)}$ is the external potential that reproduces the target density via the many-body Schr\"{o}dinger equation. This approach is general and not restricted to $|\Psi\ra$ being a ground-state.
 We numerically implement this scheme with $80\%$ mixing of the previous potential to reduce the chance of instabilities.

\subsection{Wave function and density metrics}
To describe the closeness between the wave functions (or the densities) of the exact and i-LDA systems we consider the rigorous `natural' metrics for wave functions ($D_{\psi}$) and densities ($D_{\rho}$) discussed in Ref.~\cite{DistancePRL}. For completeness we report below their expression
\begin{eqnarray}
 &&D_{\psi}(\psi_{1},\psi_{2})=\left[2N-2\abs{\int\psi_{1}^{*}\psi_{2}d\mbf{r}_{1} \ldots d\mbf{r}_{N}}\right]^{\frac{1}{2}},  \label{dpsi}\\
 &&D_{\rho}(\rho_{1},\rho_{2})=\int\abs{\rho_{1}(\mbf{r})-\rho_{2}(\mbf{r})} d\mbf{r}, \label{drho}
\end{eqnarray}
  where we have followed the convention in Ref.~\cite{DistancePRL} and normalised the wave functions to the particle number $N$.

To facilitate straightforward comparisons we use scaled metrics $\hat{D}_{\psi}$ and $\hat{D}_{\rho}$ that reside on $[0,1]$:
\begin{eqnarray}
 &&\hat{D}_{\psi}(\psi_{1},\psi_{2})=\frac{D_{\psi}(\psi_{1},\psi_{2})}{\sqrt{2N}}\\
 &&\hat{D}_{\rho}(\rho_{1},\rho_{2})=\frac{D_{\rho}(\rho_{1},\rho_{2})}{2N}.
\end{eqnarray}

 \subsection{Potential metrics} 
For a finite lattice system of $d$ sites and finite potential we may define a distance between potentials similar to the density distance created from the density norm in Ref.~\cite{DistancePRL}
\begin{equation}
\tilde{D}_{v}^{A}=\frac{1}{d}\sum_{j}^{d}|v_{1,j}-v_{2,j}|, \label{v_metric}
\end{equation}
where $v_{1,j}$ ($v_{2,j}$) is the external potential of system 1 (2) at site $j$.  As the sum of the potential is not constrained, unlike the density, then we have divided by the number of sites $d$ to allow fair comparison between different system sizes. A metric $d(x,y)$ must adhere to the following three conditions: $d(x,y)=d(y,x)$, $d(x,y)=0 \Leftrightarrow x=y$, and the triangle inequality $d(x,z)+d(z,y)\geq d(x,y)$.   We see that Eq.~\ref{v_metric} is symmetric on exchanging $v_{1}$ with $v_{2}$, is zero if and only if $v_{1}$ and $v_{2}$ are equal, and as the absolute value satisfies the triangle inequality then so does $\tilde{D}_{v}^{A}$. Hence the potentials on a lattice of $d$ sites with the distance Eq.~\ref{v_metric}  give rise to a metric space. The metric (Eq.~\ref{v_metric}) could be applied to any couple of systems whose Hamiltonians satisfy (Eq.~\ref{H}); for the scope of this paper we will use it to compare systems whose Hamiltonians differ by the external potential term only, and in particular the two external potentials are obtained by using two different methods/approximations to solve the same physical problem.
We note that the metric (Eq.~\ref{v_metric}) would not be suitable for a continuous variable system since the potential for many systems is unbounded unless a cut-off is used.

Physical potentials are only defined up to an additive constant $c$. This needs to be taken into account to prevent the unwanted situation where the wave function and density distances are both zero, but the potential distance is not. So, similarly to the wave function distance \cite{DistancePRL} that was created to be gauge-independent in Ref.~\cite{DistancePRL}, we wish to remove the arbitrarity of this constant for the potential.  We define then

\begin{equation}
D_{v}^{A}=\min_{c}\frac{1}{d}\sum_{j}^{d}|v_{1,j}-v_{2,j}+c|,
\end{equation}
and we find the $c\in \Re$ that minimizes this sum using an iterative weighted least squares technique \cite{Schlossmacher73},

\begin{equation}
D_{v}^{A}=\frac{1}{d}\sum_{j}^{d}|v_{1,j}-v_{2,j}+c_{\text{min}}|. \label{DVA}
\end{equation}

We consider Eq.~\ref{DVA} as a distance between {\em classes} of potentials such that each class contains all potentials which are equal up to constant. Eq.~\ref{DVA} 
 is symmetric on exchanging $v_1$ with $v_2$.  Also $D_{v}^{A}(v_1,v_2)=0$ if and only if $v_1$ and $v_2$ belong to the same class. Then, to demonstrate that Eq.~\ref{DVA} is a metric between physically different potentials, it only remains to show that it satisfies the triangle inequality. Now for the $c_1$ and $c_3$ that minimize the sums we have
\begin{equation}
\nonumber
D_{v}^{A}(v_1,v_2)+D_{v}^{A}(v_2,v_3)=\tilde{D}_{v}^{A}(v_1+c_1,v_2)+\tilde{D}_{v}^{A}(v_2,v_3+c_3)
\end{equation}
as $\tilde{D}_{v}^{A}$ obeys the triangle inequality then

\begin{equation}
\nonumber
\tilde{D}_{v}^{A}(v_1+c_1,v_2)+\tilde{D}_{v}^{A}(v_2,v_3+c_3) \geq \tilde{D}_{v}^{A}(v_1+c_1,v_3+c_3)
\end{equation}
and
\begin{equation}
\nonumber
\tilde{D}_{v}^{A}(v_1+c_1,v_3+c_3) \geq \min_{c} \tilde{D}_{v}^{A}(v_1+c,v_3)={D}_{v}^{A}(v_1,v_3).
\end{equation}
Combining the above equations leads to the triangle inequality
\begin{equation}
D_{v}^{A}(v_1,v_2)+D_{v}^{A}(v_2,v_3)\geq {D}_{v}^{A}(v_1,v_3).
\end{equation}

We also consider a distance between potentials similar to the wave function distance defined in \cite{DistancePRL} using the sum of squares:

\begin{equation}
D_{v}^{B}=\min_{c}\sqrt{\frac{1}{d}\sum_{j}^{d}(v_{1,j}-v_{2,j}+c)^{2}}.
\end{equation}

Without the minimization, $\tilde{D}_{v}^{B}$  is the Euclidean distance scaled by $1/\sqrt{d}$ and so is a metric for potentials on $d$ sites. The same arguments as for $D_{v}^{A}$ then can be used to show that $D_{v}^{B}$ is also a metric.

This time the minimization may be achieved analytically resulting in
\begin{eqnarray}
c_{\text{min}}&=&\frac{1}{d}\sum_{j}^{d}(v_{1,j}-v_{2,j})\\
&=&\mu(\Delta v)
\end{eqnarray}
 where $\mu$ is the mean value of $\Delta v_{j}=v_{1,j}-v_{2,j}$.  This leads to

\begin{eqnarray}
D_{v}^{B}&=&\sqrt{\frac{1}{d}\sum_{j}^{d}(\Delta v_{j}-\mu(\Delta v))^{2}}\\
&=&\sigma(\Delta v) \label{DVB}
\end{eqnarray}
which was used to quantify the match of two potential curves in Ref.~\cite{CoeJCP12}.
Here $\sigma(\Delta v)$ is the standard deviation of $\Delta v$.

Both potential distances can be scaled in a standard way to a distance between $0$ and $1$ (see, e.g., \cite{Sutherland}) 
\be
\hat{D}_{v}=\frac{D_{v}}{D_{v}+1}. \label{eq:vdist}
\ee

\section{Results and discussion}

We consider the one-dimensional Hubbard model (HM):

\begin{eqnarray}
& &H_{\text{HM}}=-t\sum_{i,\sigma}\left (\hat c_{i,\sigma}^{\dagger}\hat c_{i+1,\sigma}+\hat c_{i+1,\sigma}^{\dagger}\hat c_{i,\sigma} \right) \nonumber\\
& & +U\sum_{i}\hat{n}_{i,\uparrow}\hat{n}_{i,\downarrow}+\sum_{i,\sigma}v_i\hspace{0,1cm}\hat{n}_{i,\sigma},
\label{eqn:HubbardHamiltonian}
\end{eqnarray}
where $t$ is the hopping parameter, $U$ the on-site interaction,  and $\hat c_{i,\sigma}^{\dagger},\hat c_{i+1,\sigma}$ are creation and annihilation operators of fermionic particles with $z$-spin component $\sigma=\pm1/2$ at site $i$. The metric-space analysis will be performed for three very distinct types of external potentials: homogeneous potential, harmonic potential and localized impurities. In all calculations we will set $t=1$.   For small chains ($d\leq14$ sites) we obtain exact data via Lanczos diagonalization with tolerance $10^{-14}$, while for larger chains we produce nearly-exact results using DMRG techniques whose parameters (finite-size algorithm, basis-size 80, truncation error $10^{-5}$), for $d=10$, produced deviations smaller than $0.001\%$ from exact distances. Finally the inversion scheme Eq.~\ref{eq:schemeHub} is performed with an average site error threshold of $10^{-8}$.

\subsection{Metric-space analysis: homogeneous potential}

\begin{figure}\centering
\includegraphics[width=0.47\textwidth]{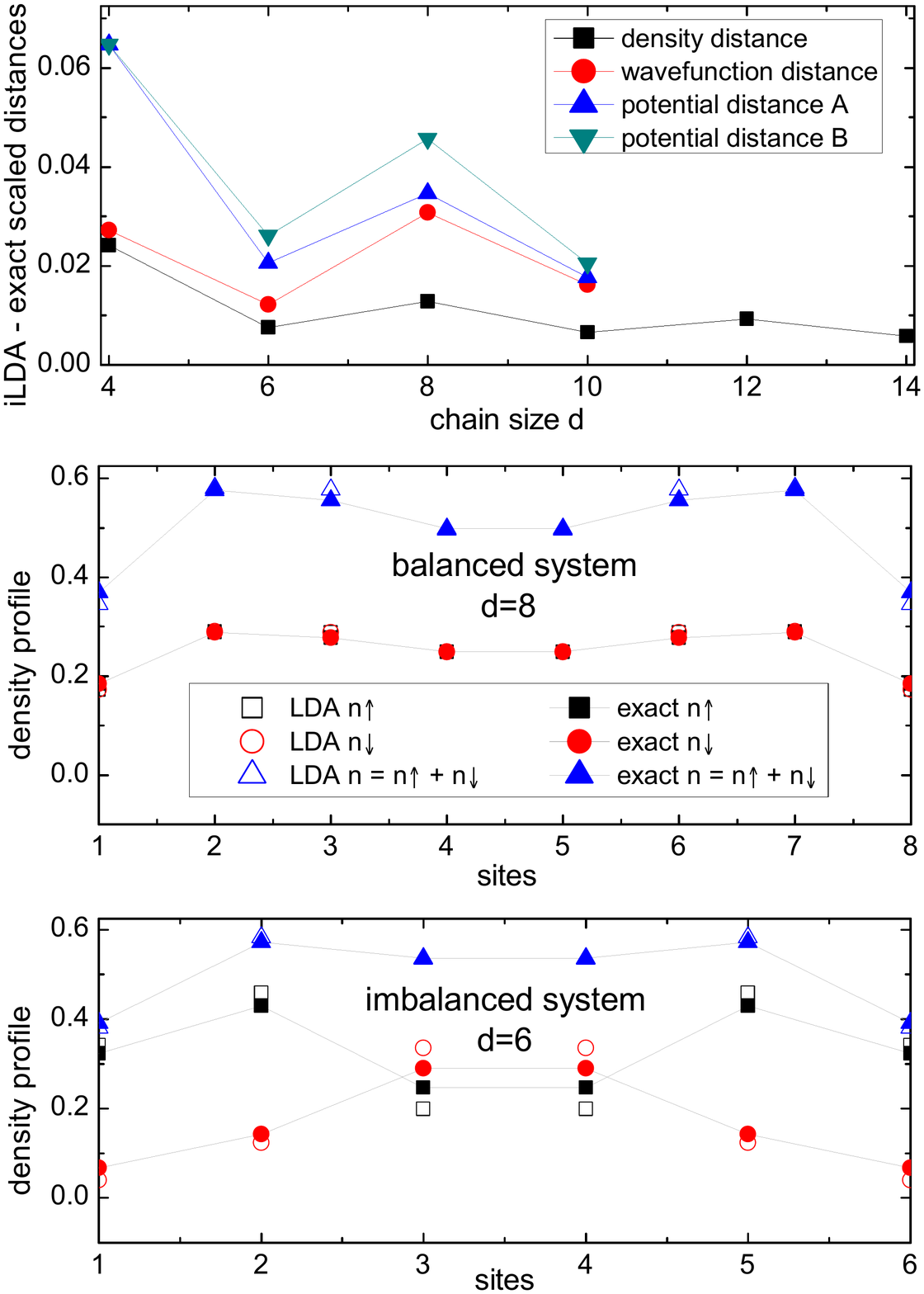}
\caption{Upper panel: Scaled distances between i-LDA and exact system for density ($\hat{D}_{\rho}$), wave function ($\hat{D}_{\psi}$)  and potential ($\hat{D}_{v}^{A}$ and $\hat{D}_{v}^{B}$) as a function of the chain size. Intermediate panel: exact and LDA density profiles for $d=8$ and N even. Lower panel: exact and LDA spin density components profiles for $d=6$ and N odd. All cases have $U=4$, $N=d/2$ and open boundary conditions.}\label{fig:distance_homog}
\end{figure}

We start by considering small Hubbard chains with no external potential,  $v_i=0$ for all $i$, so the inhomogeneity comes from finite-size effects only. In the upper panel of Figure \ref{fig:distance_homog} we show the distances  between the i-LDA potential, wave function, and density with the corresponding exact quantities for varying chain length $d$, and fixed average density  $n=N/d=0.5$, with $N$ the total number of particles. The i-LDA wave functions and i-LDA potential are obtained using the inversion scheme Eq.~\ref{eq:schemeHub}.

First we observe that, for all distances, the chains with odd number of particles are systematically closer to the exact system than the ones with $N$ even. The odd-even oscillation appears because for fixed $n$ we have alternating non-magnetic (even chains, $N_{\uparrow}=N_{\downarrow}=N/2$) and magnetic systems (odd-$N$ chains,  $N_{\uparrow}\neq N_{\downarrow}$). For non-magnetic systems, the LDA will give the same results for both spin density components, so that any error in a spin component will be doubled-up when considering the total density. For magnetic systems, spin-up and spin-down densities are different and hence the (inhomogeneous) density oscillations will be different for each component. In this case, when the two components are summed up to yield the total density, cancellation of errors in the LDA results may occur. Indeed this is what systematically happens in this system, as illustrated in the intermediate and lower panels of Fig.~\ref{fig:distance_homog}.

We also find that, with the exception of $\hat D_\psi$,  the distances for even (odd) number of particles behave qualitatively similarly: they decrease as the chain (and particle number) increases, which is consistent with the fact that the LDA becomes exact at the limit $d\rightarrow \infty$. In particular, both ways of quantifying the potential distance capture the trend. As all plotted distances are scaled to a maximum of 1, we can deduce that, for homogeneous chains and all quantities analyzed, the LDA performance is closer than $10\%$ to the exact results. However, when comparing the different quantities, we note that LDA results for densities are closer to the exact results, while external potentials are reproduced slightly less precisely by the LDA, no matter which of the potential metrics we use.

The wave function distance reproduces the odd/even N oscillation behaviour; however for chains with even (odd) number of particles the distance is increasing with number of sites. In addition the i-LDA wave function seems to perform, as the number of sites grows,  increasingly worse than the other  i-LDA quantities: it gives the overall closest distance for $d=4$ but the farthest for $d=10$; unfortunately the size of the Hilbert space becomes too big to perform the inversion scheme for larger values of $d$, so we cannot confirm the trend for longer chains. 
We attribute this worsening in performance to the scaling with $N$ and $d$ of the overall space: this only increases linearly with the sites for the other quantities, but the wave function (Hilbert) space exponentially grows from  $\binom{4}{1} \binom{4}{1}=16$ to  $\binom{10}{3} \binom{10}{2}=5400$.

 To illustrate this we consider the distances from the exact wave function and its density when using random wave functions and their associated densities. The only constraints on these random wave functions are that they are normalized, and the numbers of spin-up and spin-down particles are fixed at   $N_{\uparrow}= N_{\downarrow}$ for even $N$ and  $N_{\uparrow}= N_{\downarrow}+1$ for odd. This is achieved by assigning a pseudo-random number from $[-1,1]$ for each permissible Slater determinant then normalizing the resulting wave function. We see in Table \ref{tbl:RandWavDist} that picking a close wave function from this collection is much less likely than finding a wave function that gives a close density.  This becomes more pronounced as the number of sites increases causing the size of the wave function configuration space to dramatically enlarge. The densities actually become closer to the exact on average as the sites increase, which we speculate as due to the relative homogeneous density of the exact system, while by $10$ sites we find that a random wave function is almost always around the maximum distance from the exact.

\begin{table}[h]
\centering
\begin{tabular}{|c|c|c|}\hline
Sites & Mean  $\hat{D}_{\psi}$      & Mean $\hat{D}_{\rho}$      \\ \hline
4     & 0.889      & 0.143        \\ 
6     & 0.956     & 0.081   \\ 
8     & 0.986     &   0.070\\ 
10    & 0.995    &  0.052 \\ \hline
\end{tabular}
\caption{\label{tbl:RandWavDist} Mean scaled distances $\hat{D}_{\psi}$ and  $\hat{D}_{\rho}$  to three decimal places of one million random wave functions and their densities from the exact wave function and exact density for $U=4$, $N=d/2$ and open boundary conditions. } 
\end{table}

\subsection{Metric-space analysis: localized impurities}

Here the external potential is chosen to be a collection of localized repulsive impurities with the same strength $V$. We start by considering a chain of size $d=10$ with open boundary conditions and with two impurities localized at the central sites.

\begin{figure}
\centering
 \includegraphics[width=.47\textwidth]{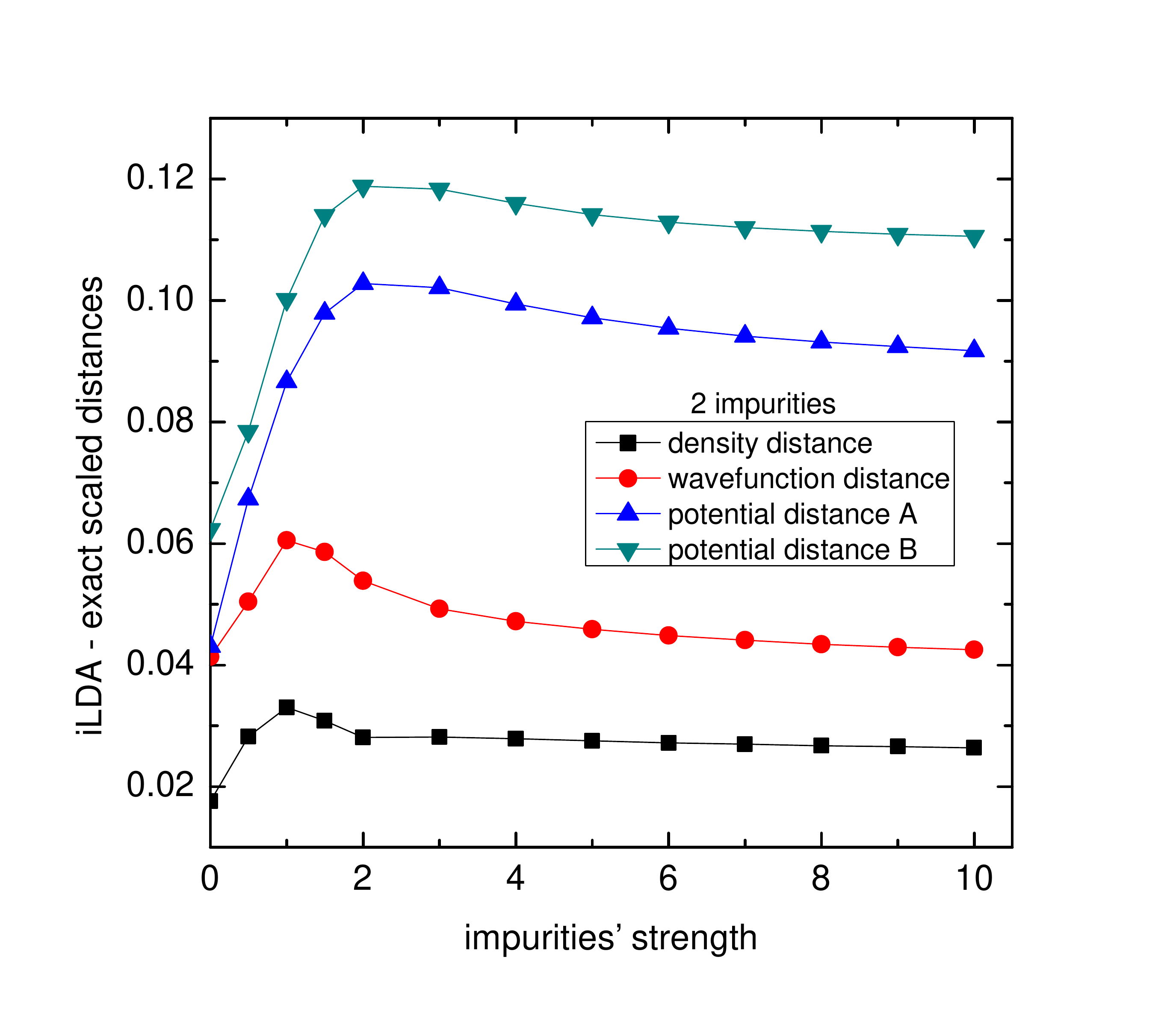}
\caption{Scaled distances ($\hat{D}_{\rho}$, $\hat{D}_{\psi}$, $\hat{D}_{v}^{A}$ and $\hat{D}_{v}^{B}$) between i-LDA and exact system as a function of the impurities' strength $V$ for chains with 2 impurities at the center. Other parameters are: $U=4$, $N=4$, $d=10$ and open boundary conditions.}\label{fig:2imp_dist}
\end{figure}

Figure \ref{fig:2imp_dist} presents the distances for density, wave function and potential between the i-LDA and the exact system as a function of the impurities strength $V$. We see that the distances have a similar qualitative behavior: they are minimum at $V=0$, show a peak for small/intermediate $V$'s, and saturate for $V>>1$.  The saturation occurs because the increasing repulsion $V$ progressively devoids the impurity sites until their occupation becomes negligible, and then the density profile at the remaining sites remains unaffected by further increase of $V$. Interestingly the peak appears at significantly higher $V$ value for the potential distances than for the wave function and density distances, which are  in general quantitatively much more similar. At contrast with Fig.~\ref{fig:distance_homog}, here we are keeping $N$, and hence  the size of the configuration space, fixed. The regime $V>>1$  can be considered as equivalent to reducing it. Instead, what strongly varies is the potential strength. This is found to affect  the potential distances differently from the other distances: for intermediate and large $V$'s, the potential distances are substantially larger than the density and wave function distances. 

\begin{figure}\centering
 \includegraphics[width=.47\textwidth]{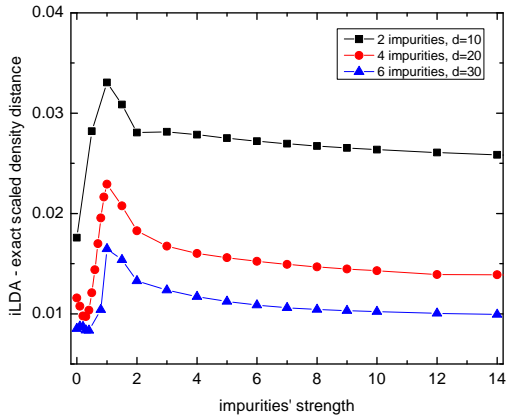}
\caption{ Scaled density distance  ($\hat{D}_{\rho}$) between the i-LDA and the exact system as a function of the impurities' strength $V$ for chains with 2, 4 and 6 impurities at the center. We keep the percentage of impurities fixed at $20\%$ of the sites, so we have $d=10$ (for 2 impurities), $d=20$ (for 4 impurities) and $d=30$ (for 6 impurities). For all cases $n=0.4$, $U=4$ and open boundary conditions. }\label{fig:density_dist}

\end{figure}

\begin{figure*}[h!]
\centering
 \includegraphics[width=1\textwidth]{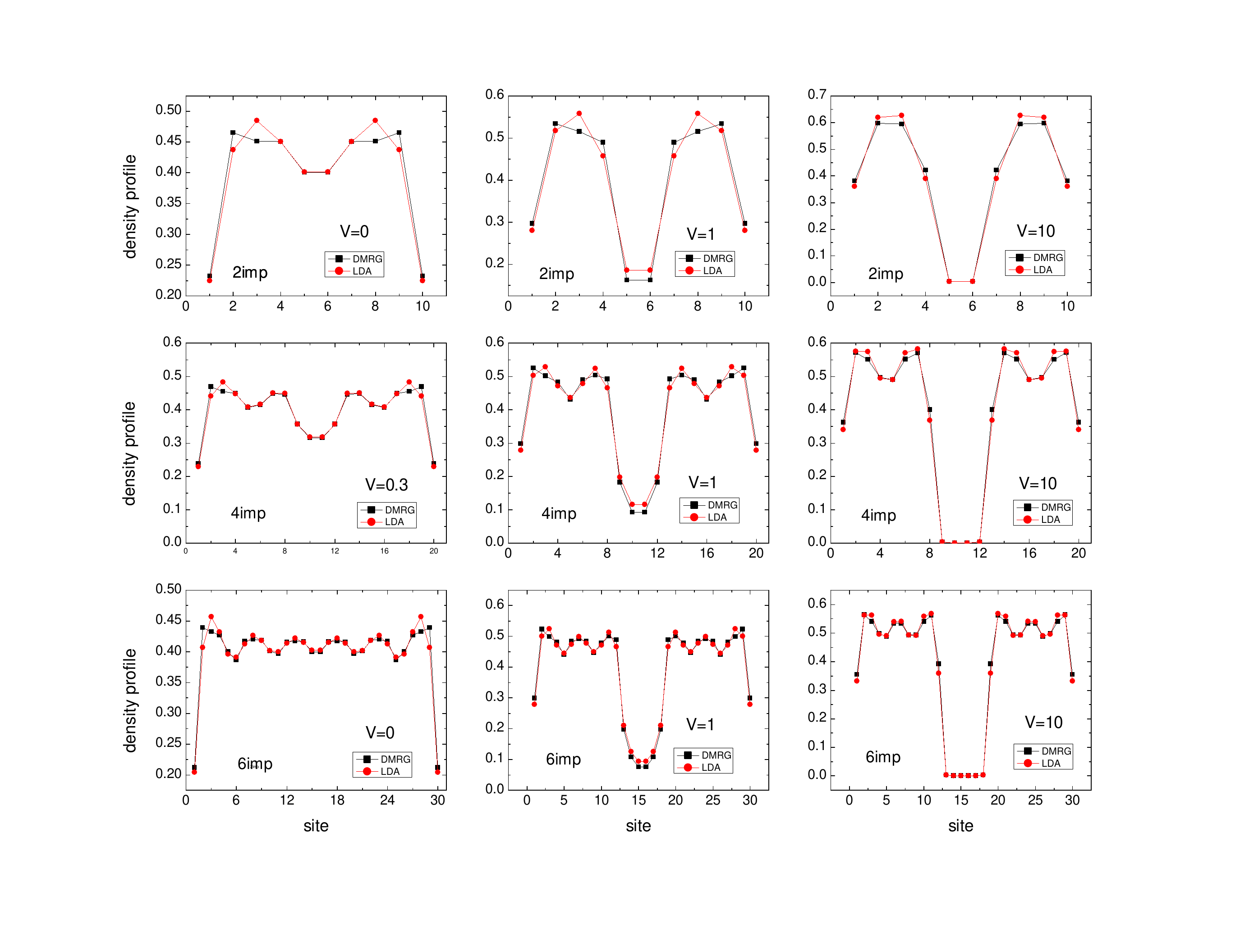}
\caption{Density profiles for specific values of $V$ of the impurities systems presented in Fig.~\ref{fig:density_dist}.}\label{fig:dens_profiles}
\end{figure*}

\begin{figure}\centering
 \includegraphics[width=.47\textwidth]{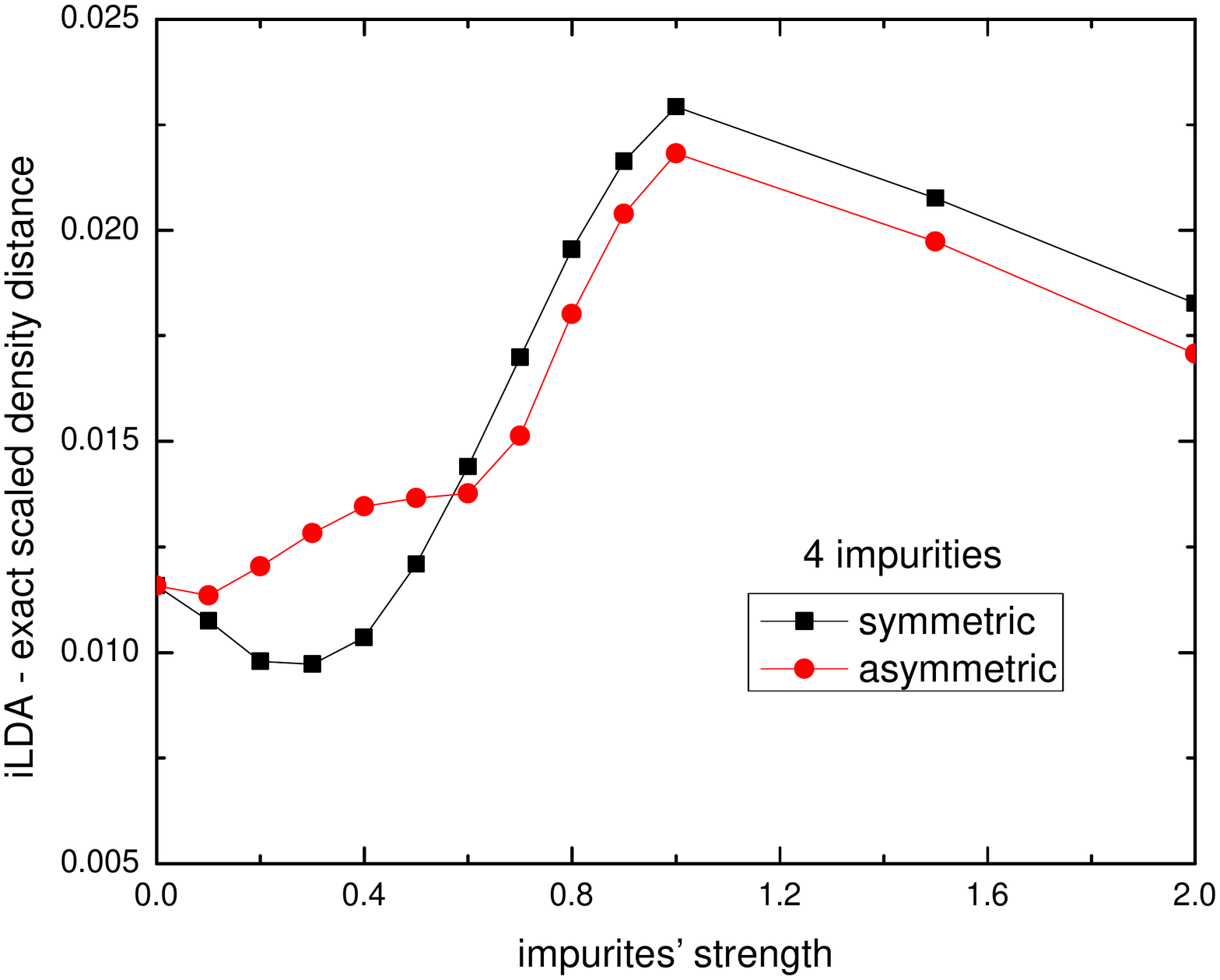}
 \includegraphics[width=.47\textwidth]{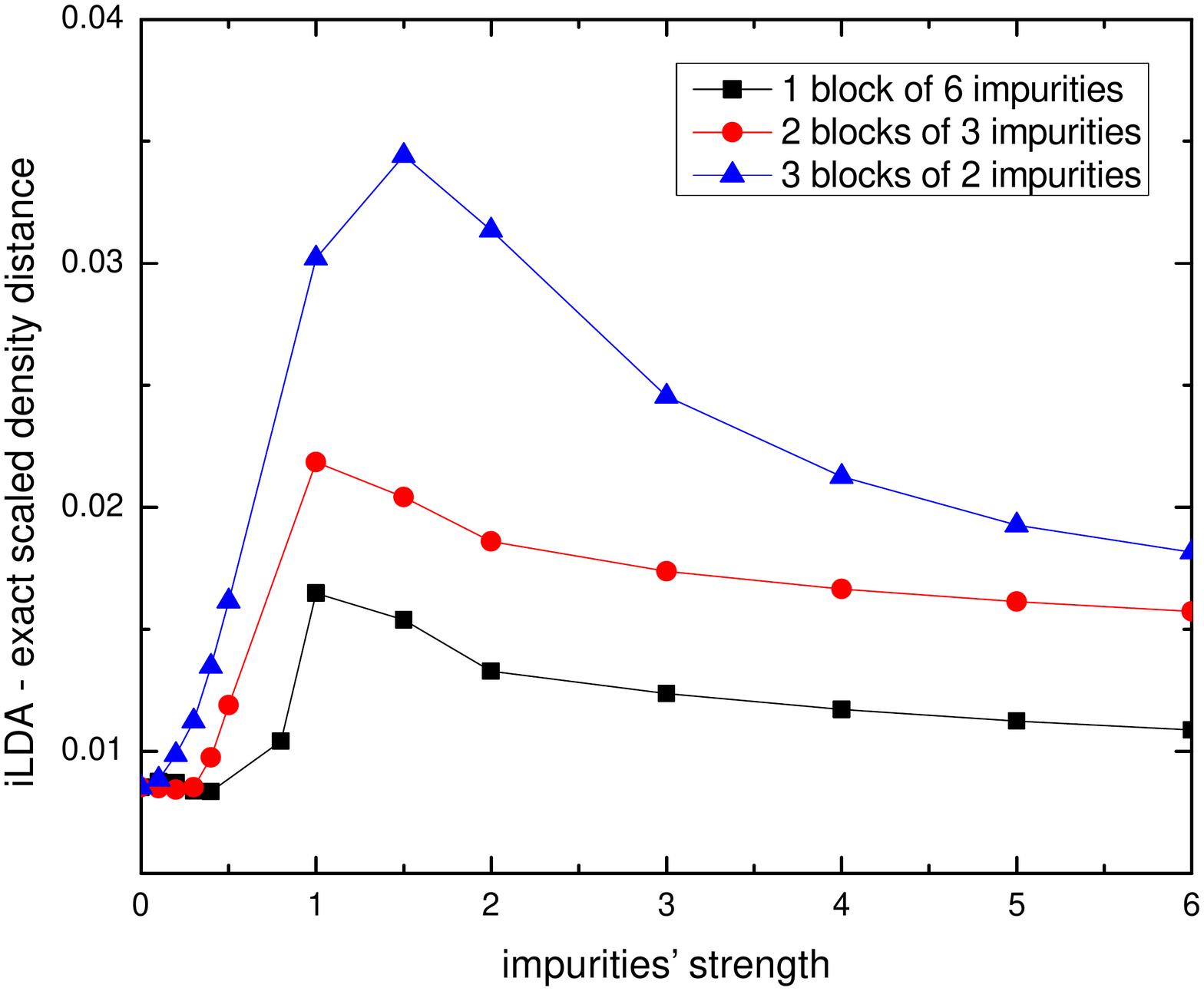}
\caption{Upper panel: Scaled density distance ($\hat{D}_{\rho}$) as a function of the impurities' strength $V$ for chains of size $d=20$, with 4 impurities symmetrically (at sites $9-12$ ) and asymmetrically located (at sites $13-16$), with $n=0.4$, $U=4$ and open boundary conditions. Lower panel: Scaled density distance  ($\hat{D}_{\rho}$) between i-LDA and exact system for chains with 6 impurities divided in blocks which were symmetrically distributed in the chain: a single block with 6 impurities (at sites $13-18$), 2 blocks with 3 impurities each (at sites $9-11$ and $20-22$) and 3 blocks with 2 impurities each (at sites $7-8$, $15-16$, $23-24$), all cases with $n=0.4$, $d=30$, $U=4$ and open boundary conditions. }\label{fig:asymetric}
\end{figure}

In the regime  $V>>1$, impurities $V$ mimic extra boundaries, and they make the LDA poorer, so one would expect this regime to correspond to the maximum distance; because of this the presence of a peak at small/intermediate $V$'s is unexpected. In order to understand its appearance,  in Figure \ref{fig:density_dist} we analyze the density distance for larger chains, $d=20$ and $d=30$, keeping the same impurity percentage and structure (centered impurities). Our results show that  not only the peak persists, but also a minimum at $V\neq 0$ appears \footnote{The presence of this minimum cannot be confirmed for $d=10$ as both LDA and DMRG results do not converge accurately for $0< V < 1$}. This is surprising because the LDA is, by definition, exact at the spatially homogeneous limit: in our finite chains one would expect that $V=0$ is the closest to this limit, as then only the boundaries induce inhomogeneities, so that $V=0$ should correspond to the minimum distance from the exact system.

We solve these conundra by analyzing the density profiles (Fig.~\ref{fig:dens_profiles})these show that both the dips and the peaks of Fig.~\ref{fig:density_dist} are generated by the finite-size of the chains considered. For small/intermediate $V$ the peak reflects the fact that the impurities sites are still non-empty, so the corresponding strong density inhomogeneity contributes to make the LDA worse in comparison to the saturation regime where the impurities sites are empty. Instead the dip observed in Fig.~\ref{fig:density_dist} is related to the particular choice of locating the impurities  symmetrically with respect to the boundaries: this symmetry somehow favors the LDA performance. When we simply displace the impurities to an asymmetric position, the dip disappears, as we can see in the upper panel of Figure \ref{fig:asymetric}.

In general we find that the LDA's performance worsens for shorter chains and same impurity structure (Fig.~\ref{fig:density_dist}), and for a symmetric but increasing spreading of the impurities with same chain length (lower panel of Fig.~\ref{fig:asymetric}).
It is easier to understand this behavior in the saturation regime $V>>1$, where impurity sites are practically empty and then the chains get fragmented into increasingly smaller segments, bearing higher inhomogeneity. For intermediate values of $V$, when the impurity sites are just depleted but not empty, the LDA performance may even worsen, leading to the peaks visible in both Fig.~\ref{fig:density_dist} and Fig.~\ref{fig:asymetric}.  This is because in this case the LDA has  also to cope with simulating the highly inhomogeneous impurity density dips (see panels with $V=1$ in Fig.~\ref{fig:dens_profiles}).

\begin{figure}\centering
 \includegraphics[width=.5\textwidth]{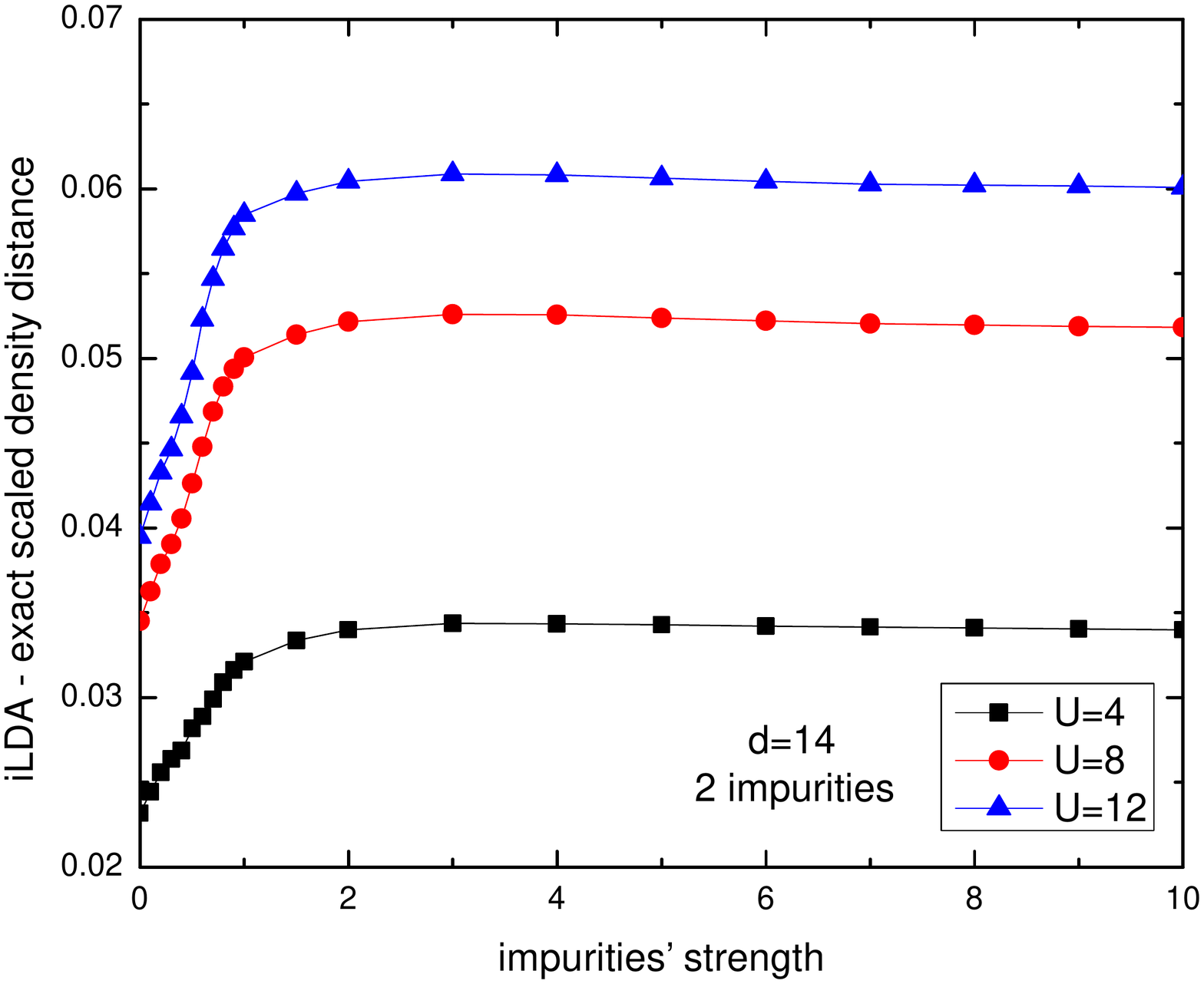}
\caption{
Scaled density distance  ($\hat{D}_{\rho}$) between i-LDA and exact system as a function of impurities' strength $V$ for chains with 2 impurities at the center and different interaction strengths $U$. Other parameters are: $d=14$, $N=4$ and open boundary conditions. }\label{fig:L14_severalU}
\end{figure}

Finally in Figure \ref{fig:L14_severalU} we consider the LDA density distances  with respect to the impurity strength for  distinct values of the interaction $U$ and fixed chain length ($d=14$). Impurities here are in smaller concentration with respect to Fig.~\ref{fig:density_dist} but still located symmetrically in the  center of the chains, so maximizing homogeneous {\it segments} length with respect to the chain size. We find that the density distance properly describes the expected behavior that the LDA performs worse for higher interaction $U$. The results also show how sensitive the LDA is to finite-size effects: a symmetric but smaller concentration of impurities is here enough to destroy the dip-and-peak structure.

\subsection{Metric-space analysis: harmonic potential}

 We now turn to consider systems confined by the harmonic potential $kx^{2}$, centered between the two middle sites. We use open boundary conditions, $d=8$ sites, interaction strength $U=2$ and $N=2$ particles. We analyze the LDA performance using the distance between the i-LDA and the exact systems as a function of $k$ for potential,  wave function and density.

\begin{figure}
\centering
 \includegraphics[width=.4\textwidth]{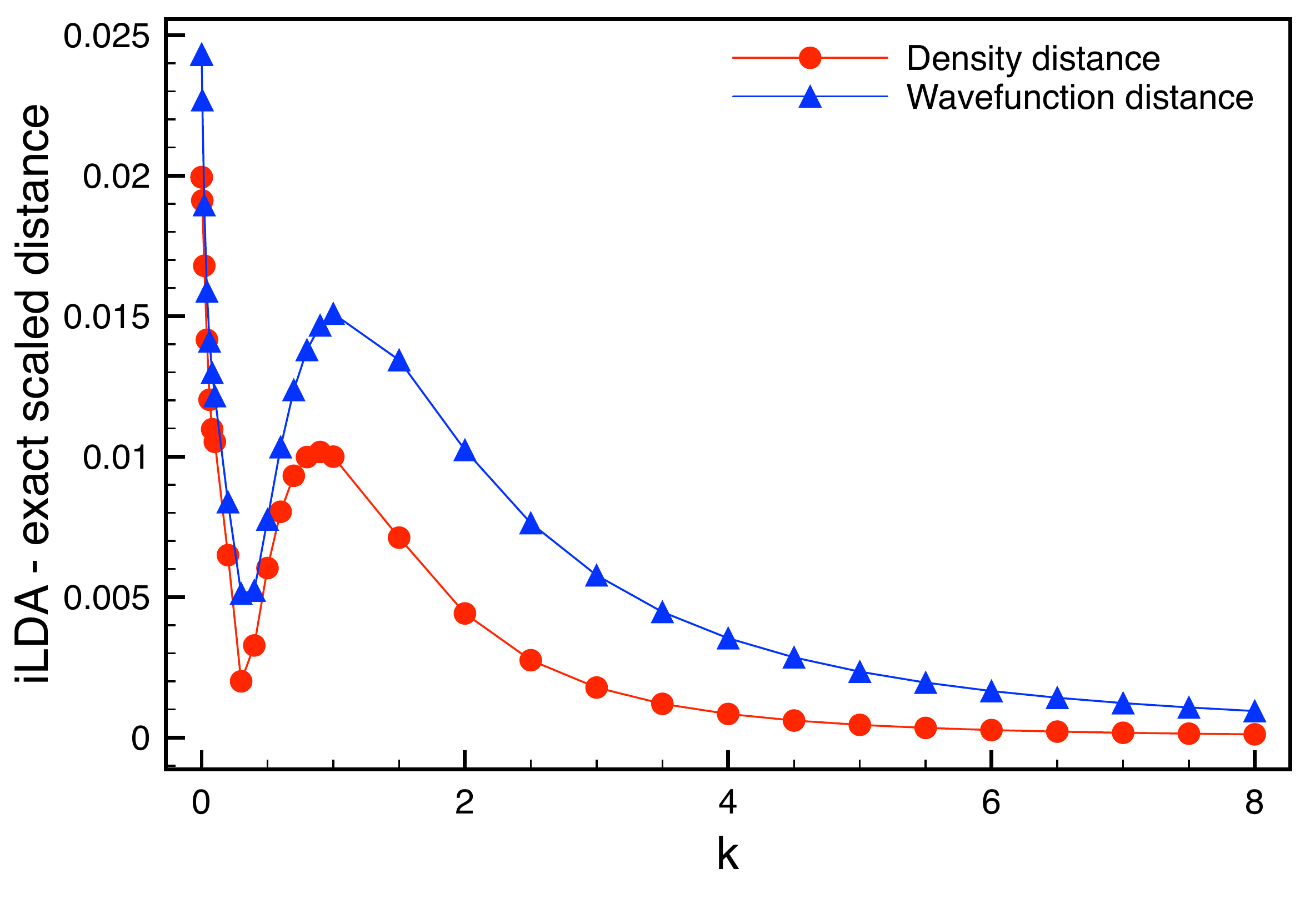}
  \includegraphics[width=0.4\textwidth]{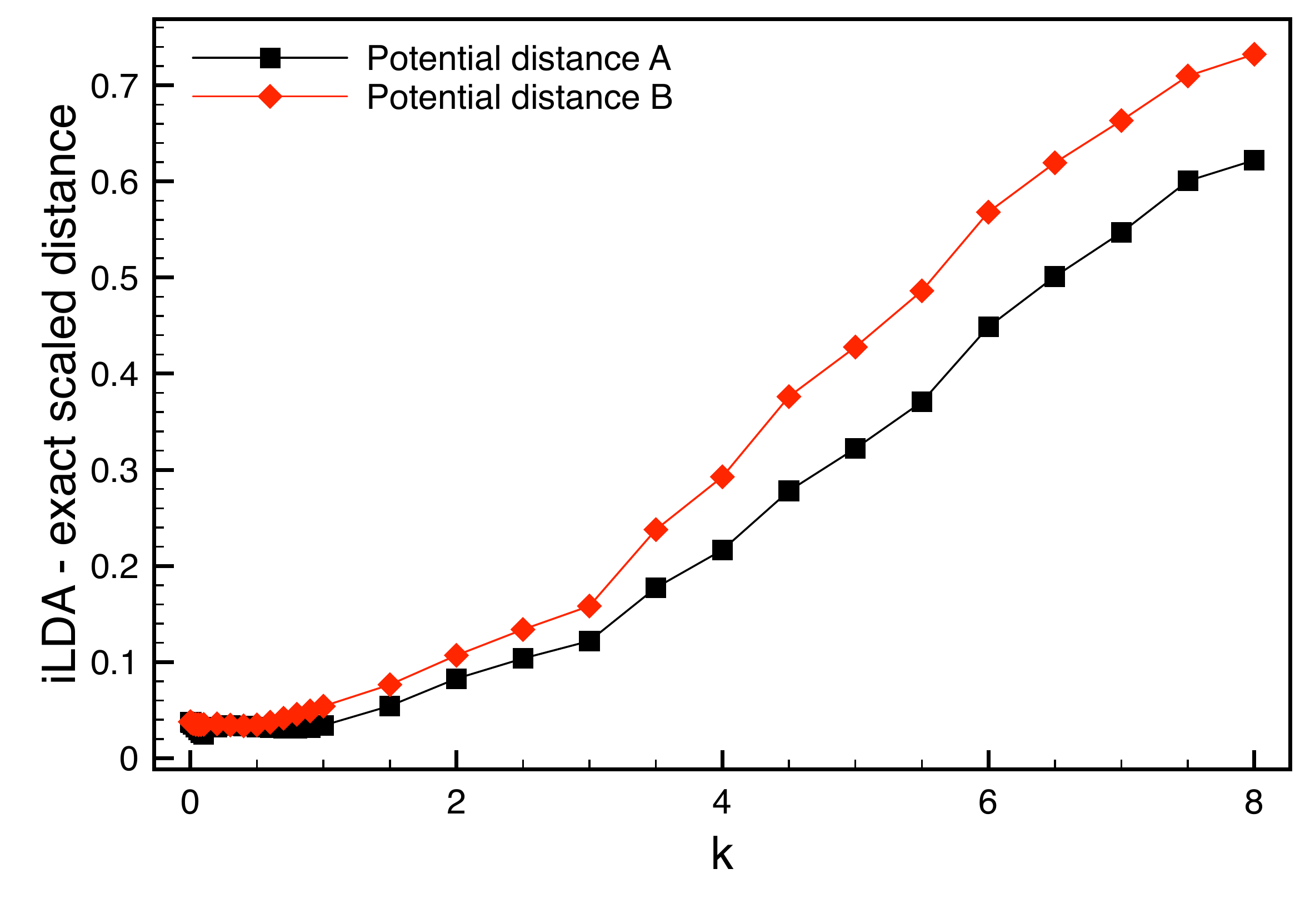}
\caption{Upper panel: Scaled density ($\hat{D}_{\rho}$) and wave function ($\hat{D}_{\psi}$) distances between i-LDA and exact systems  versus confining potential strength $k$. Lower panel: Scaled potential distances ($\hat{D}_{v}^{A}$ and $\hat{D}_{v}^{B}$)   between i-LDA and exact systems, versus confining potential strength $k$. In both panels: $U=2$, $d=8$ and $N=2$, with confining potential $kx^{2}$ symmetric with respect to chain centre.}\label{fig:DenAndWavHarmonic}
\end{figure}

Density and wave function distances are considered in the upper panel of Fig.~\ref{fig:DenAndWavHarmonic}:  the LDA performance is very good for all $k$'s, with less than 3\% maximum error for $k\to 0$. Also, as for the localized impurity case, density and wave function distances behave very similarly: they both present a  local minimum at $k\approx 0.3$, a local maximum around $k=1$, and tend to zero as the confining potential increases further.
We attribute this minimum-maximum structure to a competition between depletion of density in certain sites and effective reduction of the chain length, similarly to the mechanism we have described for Fig.~\ref{fig:density_dist}. In the present case, as $k$ starts to increase, the parabolic potential starts to deplete the chain ends' sites, excluding more sites as the potential strengthens. For large $k$ values, all central sites become double occupied, the system freezes, and the distances between densities and between wave functions tend to zero.

Conversely, both potential distances display a behavior {\it completely different} from wave functions and densities (lower panel of Fig.~\ref{fig:DenAndWavHarmonic}): they are monotonically increasing with $k$ after $k=0.8$ with the minimum of $\hat D_{v}^{A}$ occurring around $k=0.1$ while that of $\hat D_{v}^{B}$ is at $k\approx 0.4$.
Notably the LDA performance, when measured in this way, worsens as $k$ increases, up to more than 70\% error. Also both potential distances do not display any obvious sensitivity to the freezing of the system: indeed the potential can arbitrarily increase with increasing $k$, even after saturation of the central sites (freezing) has occurred. In Ref.~\cite{EPL:2015} we discussed how, for DFT on a lattice, in the limiting cases of full-filling and/or potential tending to infinity, the one-to-one correspondence between density and external potential does {\it not} hold. The large-$k$ case at hand comes close to these limiting behaviors, and this explains why the trend of the potential distances  can become strikingly different $-$ and {\em even opposite}, like in this case $-$ with respect to the behavior of densities and wavefuctions distances.

 \section{Conclusions}
Within the metric approach to quantum mechanics~\cite{DistancePRL,Sharp:2014}, we used wave function, density and external potential distances to explore LDA performances for short and medium inhomogeneous one-dimensional Hubbard chains. 

The analysis has been carried out by comparing the distance between exact many-body systems and their corresponding interacting-LDA systems, that is the interacting systems built to have the LDA ground-state densities. To derive the Hamiltonian of the interacting-LDA systems, we have  applied the inversion scheme of Ref.~\cite{EPL:2015} to the LDA density of the one-dimensional Hubbard model, and obtained, to a high degree of accuracy, the potential of the corresponding interacting system.

We considered open boundary conditions and three distinct potentials: homogeneous, localized impurities, and harmonic confinement, for different chains' length, particle numbers and interaction  strength.

The homogeneous and the localized-impurities analyses revealed several features about the LDA performance across different regimes of parameters. In particular we found interesting finite-size effects, with some counterintuitive behaviour by the LDA. Remarkably, our distances for wave functions and particle densities were sensitive to all of these effects: they correctly pointed out the regimes where the LDA performs fairly well and the regimes where it fails, as confirmed by the direct analysis of the systems' densities. Thus these metrics were proved to be useful for testing the pitfalls of the LDA and therefore could be a practical tool for testing the efficiency of any other density-functional approximation.  In addition our work provided evidence for lattice systems that when the LDA density is close to its exact counterpart then so too is the interacting LDA wave function and vice-versa. Although this relationship was affected by the size of the wave function's configuration space it was not strongly altered. We also found that the density distance is in general smaller than the others.

One important result from this work is that care must be taken when considering the distance between external potentials in appraising the performance of a density-functional approximation. At least for systems on a finite lattice, the fact that the one-to-one correspondence between density and potential fails in certain limits \cite{EPL:2015} seems to imply that the potential distance between exact and approximated systems may be misleading, even showing a behavior which is qualitatively opposite to the density and wave function distances for systems close to these limits. In these parameter regions, the potential distance suggests that the LDA behaves increasing poorly, failing to recognize that instead its performance is increasingly improving.

\acknowledgements

VVF is financially supported by FAPESP (Grant: 2013/15982-3) and CNPq (Grant: 448220/2014-8). IDA acknowledges support from CNPq (Grant: PVE - Processo: 401414/2014-0) and from the Royal Society through the Newton Advanced Fellowship scheme (Grant no. NA140436).

\end{document}